\documentclass[epj]{svjour}
\usepackage{graphicx}
\usepackage{bm}
\usepackage{hyperref}

\newcommand{\nc}{\newcommand}
\nc{\us}{U(1)$_S$}
\nc{\ph}{\phi}

\begin{document}


\title{Radiative Events as a Probe of  Dark Forces
at GeV--Scale {\boldmath $e^+ e^-$} Colliders}

\author{L. Barz\`e\inst{1}, G. Balossini\inst{2}, C. Bignamini\inst{1}, 
C.M. Carloni Calame\inst{3}, G. Montagna\inst{1}, 
O. Nicrosini\inst{2} and F. Piccinini\inst{2}}

\institute{Dipartimento di Fisica Nucleare e Teorica, Universit\`a di Pavia, Via A. Bassi 6, 27100, Pavia, Italy and\\
INFN, Sezione di Pavia, Via A. Bassi 6, 27100, Pavia, Italy \and
INFN, Sezione di Pavia, Via A. Bassi 6, 27100, Pavia, Italy \and 
School of Physics \& Astronomy, University of Southampton, 
Southampton SO17 1BJ, U.K.}

\date{\today}

\abstract{
High-luminosity $e^+ e^-$ colliders at the GeV scale (flavor factories) have been recently recognized to be an ideal environment to search for a light weakly coupled vector boson $U$ (dark photon) emerging in several new physics models. At flavor factories a particularly clean channel is the production of the $U$ boson in association with a photon, followed by the decay of the $U$ boson into lepton pairs. Beyond the
approximations addressed in previous works, we perform an exact lowest order calculation of the signal and background processes of this channel. We also
include the effect of initial and final state QED corrections neglected so far, to show how they affect the 
distributions of experimental interest. We present new results for the expected statistical significance to a dark photon signal 
at KLOE/KLOE-2  and future super-$B$ factories. The calculation
is implemented in a new release of the event generator BabaYaga@NLO, which is available for full event
simulations and data analysis.
}
\titlerunning{Radiative Events as a Probe of  Dark Forces
at GeV--Scale {\boldmath $e^+ e^-$} Colliders}
\authorrunning{L. Barz\`e et al.}
\maketitle

\section{Introduction}
\label{sect1}
In recent years, striking astrophysical observations have failed to find an interpretation in terms of standard
astrophysical or particle physics sources. Among these observations, there are the 511 KeV gamma-ray signal from the 
galactic center observed by the INTEGRAL satellite \cite{Jean:2003ci}, the excess in the cosmic ray
positrons reported by PAMELA \cite{Adriani:2008zr}, the total electron
and positron flux measured by ATIC \cite{:2008zzr}, Fermi \cite{Abdo:2009zk},
and HESS \cite{Collaboration:2008aaa}, the annual modulation of the 
DAMA/LIBRA signal \cite{Bernabei:2008yi} and the features of the low-energy spectrum of rare events 
reported by the CoGeNT collaboration \cite{Aalseth:2010vx,Aalseth:2008rx}. These evidences can
be comprehensively interpreted by well motivated extensions of the Standard Model (SM) that predict the existence of a WIMP Dark Matter 
(DM) particle belonging to a secluded gauge sector under which the SM particles are uncharged 
\cite{Boehm:2003hm,Pospelov:2007mp,Pospelov:2008jd,ArkaniHamed:2008qn,Cholis:2008wq,Mambrini:2010dq}. In particular, 
an abelian gauge symmetry, with an associated $U$ boson (``dark photon"), 
can communicate with the SM through a kinetic mixing term of the form \cite{Boehm:2003hm,Pospelov:2007mp,Holdom:1985ag}
\begin{equation}
{\cal L}_{\rm mix} \, = \, -\frac{\epsilon}{2} \, F_{\mu\nu}^{\rm em} F^{\mu\nu}_{\rm dark} 
\label{eq:fdark}
\end{equation}
where  $\epsilon$ is a kinetic mixing parameter. Annihilation of DM into the
$U$ boson, which decays into charged leptons and is light enough to kinematically
forbid a decay that produces antiprotons, can explain the electron and/or positron excesses, and 
the absence of a similar effect in the PAMELA antiproton observations. Independently of connections with DM physics, general theoretical arguments 
\cite{ArkaniHamed:2008qn,Dienes:1996zr} suggest that the kinetic mixing parameter $\epsilon$ must be naturally of the order of $10^{-4}-10^{-2}$, if the mixing occurs through loop effects. 
On the other hand  the phenomenological condition $M_U$ below the GeV scale, driven by the astrophysical observations described above, requires that $\epsilon$ is in the same range 
in order to avoid contradiction 
with the available data, in particular with the constraints imposed by the precision measurements of the 
anomalous magnetic moment of the leptons and of the electromagnetic coupling constant \cite{Pospelov:2008zw}.

From a phenomenological perspective, the motivations just discussed are, perhaps, the most compelling towards postulating the existence of a new gauge boson $U$. It is however worth mentioning that there is a broad class of new physics models that predict the existence of such a particle, without relying on DM arguments (see e.g. Refs.~\cite{Holdom:1985ag,Dienes:1996zr}). 

An intriguing consequence of the above ideas is that such a light $U$ boson (if it exists) can be directly produced in 
 a controlled environment, such as fixed target experiments \cite{Bjorken:2009mm,Freytsis:2009bh,Essig:2010xa}
 or high-luminosity $e^+ e^-$ colliders at the GeV scale (flavor factories) \cite{Borodatchenkova:2005ct,Batell:2009yf,Yin:2009mc,Bossi:2009uw,Essig:2009nc,Reece:2009un,Baumgart:2009tn,Li:2009wz}. 
 The status and perspectives of these experiments are the topics studied in the SLAC workshop ``Searches for
 New Forces at the GeV-scale'' \cite{slac}. At flavor factories, e.g. at DA$\Phi$NE, BESIII and present and future $B$-factories, a particularly clean and simple channel, which 
 is insensitive to the details of the Higgs sector of the secluded group, is the associated production of a $U$ boson and a photon, with decay of the $U$ into lepton pairs 
 \cite{Borodatchenkova:2005ct,Yin:2009mc,Reece:2009un,Li:2009wz}. A distinctive feature of
 the expected signal is the appearance of a Breit-Wigner peak in the shape of the invariant mass distribution of the
 lepton pairs induced by the mechanism of photon radiative return and corresponding to $U$ boson resonant production. The
 drawback of this channel is the fairly small value, over a wide range of the parameters, of the signal cross section in comparison with the rate of the large physics backgrounds given by the QED radiative processes $e^+ e^- \to l^+ l^- \gamma$, $l= e, \mu$, which can be rejected by cutting on the invariant mass of the lepton pair. The studies present in the literature agree on the conclusion that 
the $U\gamma$ production process allows to reach a sensitivity to the kinetic mixing parameter in the range $\epsilon \sim 10^{-3} - 10^{-2}$ at present flavor factories, for a $U$ boson with mass  $M_U $ up to a few GeV. These analyses are generally based on the evaluation of the number of signal events through the calculation of the differential cross section of the $2 \to 2$ process $e^+ e^- \to U \gamma$, including the decay of the on-shell $U$ boson into lepton pairs by means of branching ratios 
\cite{Borodatchenkova:2005ct,Reece:2009un,Li:2009wz}, and/or  an approximate estimate of the backgrounds. More importantly, 
all the studies so far performed neglect the contribution of higher order initial and final state QED corrections, which are known to be a phenomenologically relevant effect at GeV-scale $e^+ e^-$ colliders 
\cite{Actis:2010gg}.

The aim of the paper is detailed in the following. 
First, we perform an exact tree level calculation of the signal and background processes contributing to the signatures  $e^+ e^-  \to \mu^+ \mu^- \gamma, e^+ e^- \gamma$. 
Second,  we compute the effect of the most important higher order corrections induced by multiple photon radiation and vacuum polarization, that are a source of relevant systematic effects. Third, we use our calculation in order to assess its impact on the experimental sensitivity as evaluated in the literature, and show how this can be enhanced by means of event selections not considered so far. Last but not least,  
 the calculation is made available in the form of a Monte Carlo generator, which is still 
missing for such studies \cite{Bossi:2009uw,slac} and can be used for data analysis at flavor factories to study any physical observable, i.e. invariant mass distributions but also angular distributions, correlations and so on. 

The paper is organized as follows. In Section~\ref{sect2} we briefly review the  model under consideration and describe the details of our calculation. We also present a first sample of numerical results, showing in particular how the radiative corrections 
significantly distort the
shape of the invariant mass distribution which is the most important observable for experimental searches. In Section~\ref{sect3} 
 we give new results for the statistical significance to a dark photon signal at KLOE/KLOE-2 and a super-$B$ factory, to 
 complete the picture existing in the literature for the experimental sensitivity of flavor factories. In Section~\ref{sect4} we draw our conclusions.

\begin{figure}
\begin{center}
\resizebox{0.5\textwidth}{!}{%
\includegraphics{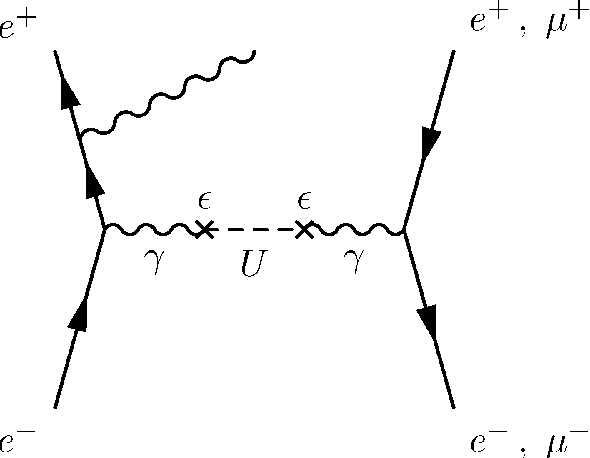}\hskip 12pt\includegraphics{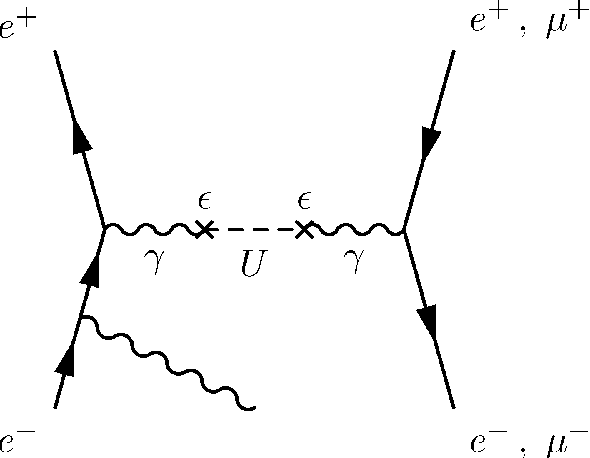}
}
\caption{Examples of Feynman diagrams with dark photon exchange contributing to the process 
$e^+ e^- \to \gamma, U \to l^+l^-\gamma$, $l=e,\mu$. The total number of diagrams is eight for the $\mu$ final state, sixteen for the $e$ final state, where also $t$-channel contributions exist. }
\label{fig:fig1}
\end{center}
\end{figure}

\section{Theoretical Framework}
\label{sect2}
Following most of the phenomenological works in the literature, we consider a minimal implementation of a secluded 
\us\ sector \cite{Batell:2009yf}. The SM Lagrangian is modified by the inclusion of a \us\ gauge group which contains DM fields, a vector 
gauge field $A^{\prime}_\mu$ and a single complex scalar Higgs field $\phi$ responsible for spontaneous symmetry breaking. The 
DM candidates are assumed to be heavy compared with the vector and Higgs bosons and the SM particles are uncharged under this 
new gauge sector. Hence, all the interactions with the SM are mediated by kinetic mixing of \us\ with the photon, if we neglect, as
usually done, the mixing with the $Z$ boson for the processes of interest here. The Lagrangian can then be written as
\begin{equation}
{\cal L}=-\frac{1}{4} F_{\mu\nu}^{\rm dark} F^{\mu\nu,{\rm dark}}  -\frac{\epsilon}{2}\, F_{\mu\nu}^{\rm dark} F^{\mu\nu} + |D_\mu \phi |^2 -V(\phi)
\label{eq:darkl}
\end{equation}
where $F_{\mu\nu}$ is the photon field strength,  $F_{\mu\nu}^{\rm dark}$ is the \us\ (dark photon) field strength, and the covariant derivative is $D_\mu=\partial_\mu+i e' A^{\prime}_\mu$ with \us\ charge $e'$. The
Higgs potential $V(\phi)$ is assumed to be of a form which spontaneously breaks the \us\ symmetry, with a vev
$\langle \phi \rangle =  v'/\sqrt{2}$. After spontaneous symmetry breaking, the  \us\  boson acquires a mass $M_U$ given by 
 $M_U = e'v'$ in the unitary gauge. As noticed in the literature  \cite{Reece:2009un}, one can remove the kinetic mixing term of 
 Eq.~(\ref{eq:fdark}) by redefining the photon $A_\mu\rightarrow A_\mu-\epsilon A'_\mu$, so that the SM fermions pick up a small U(1)$_S$
charge $\sim \epsilon e$, or, alternatively, one can treat the kinetic mixing term as an interaction as long as $\epsilon$ is small.

\begin{figure}
\begin{center}
\resizebox{0.5\textwidth}{!}{%
\includegraphics{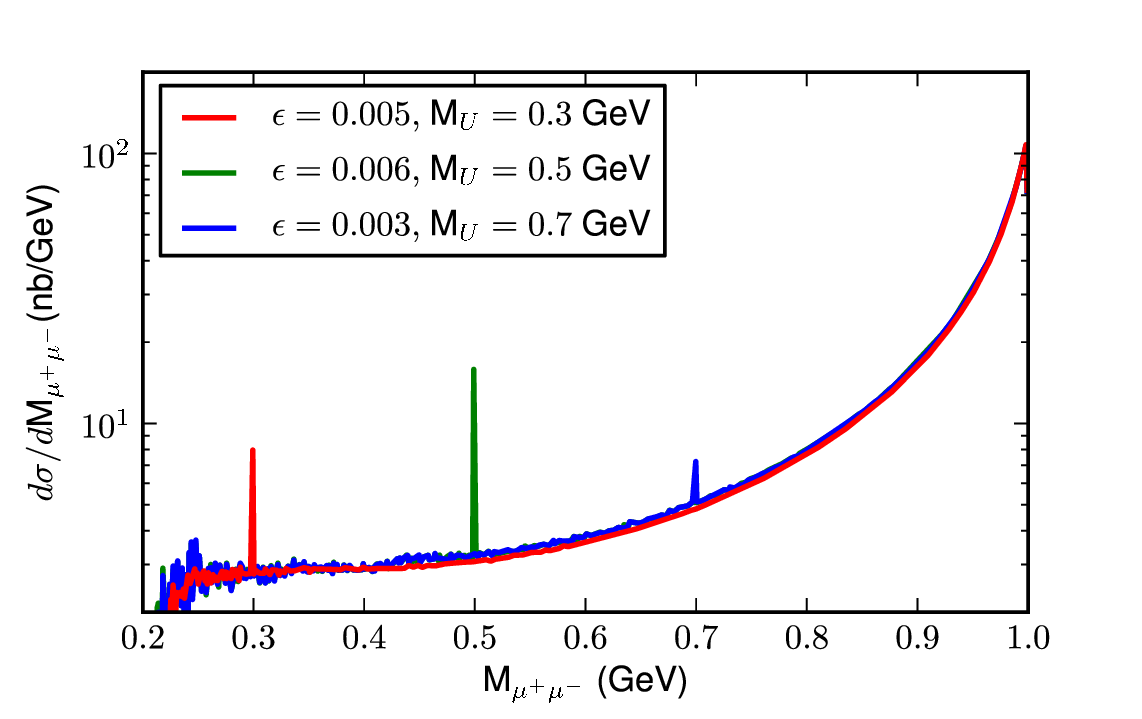}
}
\caption{Invariant mass distribution of the muon pairs for different values of the kinetic mixing parameter and 
$U$ boson mass at a $\Phi$-factory, with $\sqrt{s} = 1.02$~GeV. Predictions at the tree level. }
\label{fig:fig2}
\end{center}
\end{figure}

\begin{figure}
\begin{center}
\resizebox{0.5\textwidth}{!}{%
\includegraphics{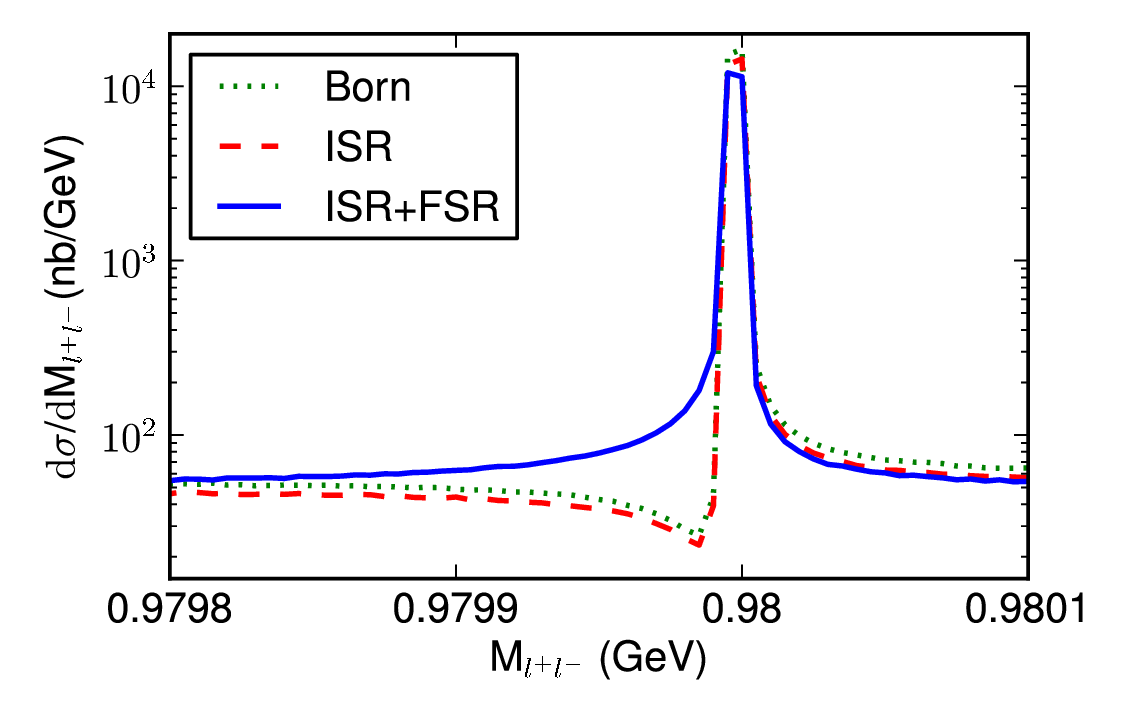}
}
\caption{The invariant mass distribution  of  muon pairs, for standard cuts at DA$\Phi$NE ($\sqrt{s} = 1.02$~GeV), and dark photon parameters $M_U = 0.98$~GeV, $\epsilon = 1 \times 10^{-3}$. Results are shown at the lowest order (dotted line), including ISR only (dashed line) and ISR + FSR (solid line). }
\label{fig:fig3}
\end{center}
\end{figure}

\begin{figure}
\begin{center}
\resizebox{0.5\textwidth}{!}{%
\includegraphics{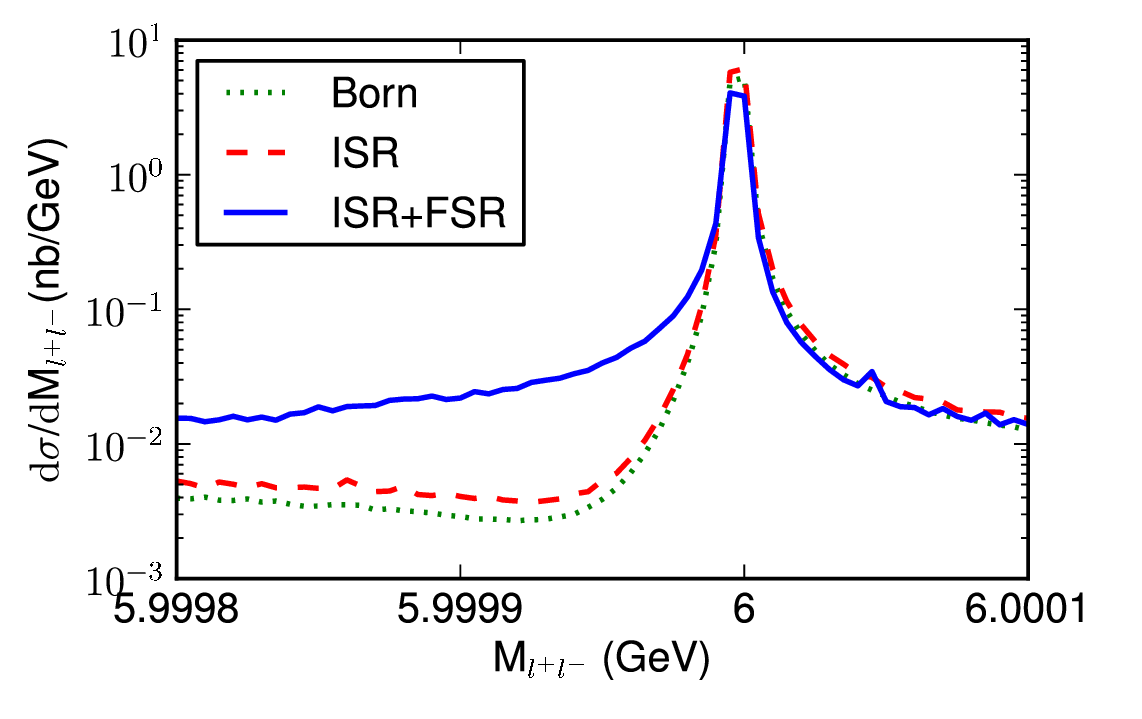}
}
\caption{The same as Fig. 3 at a $B$/super-$B$ factory ($\sqrt{s} = 10$~GeV), and dark photon parameters $M_U = 6$~GeV, $\epsilon = 5 \times 10^{-3}$. }
\label{fig:fig4}
\end{center}
\end{figure}

With this model at hand, we performed an exact lowest order calculation of the 
radiative processes  $e^+ e^-  \to \mu^+ \mu^- \gamma, e^+ e^- \gamma$, considering all possible 
$s$ and $t$-channel $\gamma$ and $U$ boson exchanges and one-photon emission from any leptons (see Fig.~\ref{fig:fig1}). Therefore, at a difference with respect to earlier works, e.g. Ref.~\cite{Reece:2009un}, the present calculation includes (i) finite width effects for $s$-channel annihilation subprocesses, both for muon and electron final states and (ii) non-resonant $t$-channel $U$ boson exchange and $s$-$t$ interference contributions for electron final states. As far as the former ones are concerned, they are of the order of $\Gamma_U / M_U$ on the integrated cross section and as such fairly small for the resonance of interest here. However, the finite width effects are a crucial ingredient from a 
phenomenological point of view because they allow, at a variance of an on-shell 
calculation, a direct evaluation of the differential cross section of main experimental 
concern, i.e. the invariant mass distribution (see Fig.~\ref{fig:fig2}) in the presence of 
arbitrary cuts, as well as an exclusive generation of the momenta of the particles 
coming from the $U$ boson decay. Moreover they are crucial in order to reconstruct the correct shape of the resonance, in particular the dip induced by $\gamma(s)-U(s)$ interference, as shown in Figs.~\ref{fig:fig3} and \ref{fig:fig4}. 
Concerning $t$-channel contributions, we checked through a number of numerical experiments that their 
impact is rather moderate, being at the per mille level or 
below it for typical $\epsilon$ values in the $10^{-3}$ range, in agreement with the
naive expectation of its order of magnitude. 
Anyway, they can in principle be useful to analyze non-resonant $U$ boson production in the $t$-channel by studying differential distributions such as angular distributions or asymmetries in the electron channel, that can be exploited in principle as a complementary search strategy, whenever a future flavor factory will be able to collect enough statistics. 
In our approach the pure $U$ boson signal contribution turns out to be defined as the difference between the full and background  matrix elements. The calculation has been done using and adapting (to $U$ boson couplings) the algorithm 
ALPHA \cite{Caravaglios:1995cd}, which is an efficient tool to compute tree level matrix elements
involving many Feynman diagrams. We implemented it in the Monte Carlo generator 
BabaYaga@NLO \cite{CarloniCalame:2000pz,CarloniCalame:2001ny,CarloniCalame:2003yt,Balossini:2006wc,Balossini:2008xr}, a standard tool for luminosity measurement at flavor factories 
\cite{Actis:2010gg}, to simulate distributions of experimental interest and account for realistic event selection criteria. 
In the $U$ boson propagator we include the total dark photon width using the formulae of Ref. \cite{Batell:2009yf} for the partial widths into leptons and hadrons, i.e. 
\begin{equation}
\Gamma_{ U \rightarrow l^+l^- }=\frac{1}{3} \alpha \epsilon^2 \, M_U \, \beta_l
\left(1+\frac{2 m_l^2}{M_U^2}\right) 
\label{eq:gammal}
\end{equation}
and
\begin{eqnarray}
\Gamma_{U \rightarrow {\rm hadrons} }=\frac{\alpha}{3} \epsilon^2 M_U \beta_l
 \left(1+\frac{2 m_\mu^2}{M_U^2}\right) R(s= M_U^2)
 \label{eq:gammah}
\end{eqnarray}
where $ R= \sigma_{e^+ e^- \rightarrow {\rm hadrons}}/ \sigma_{e^+ e^- \rightarrow \mu^+ \mu^-}$ is taken into account according to the compilation of Ref.~\cite{Teubner:2010ah} and $\beta_l = \sqrt{1-\frac{4 m_l^2}{M_U^2} }$  . It is worth noting that the $U$ boson is a dramatically tiny resonance, its total width being in the range 
$\Gamma_U \sim 10^{-5} - 10^{-2}$~MeV and $\Gamma_U \sim 10^{-7} - 10^{-4}$~MeV for $\epsilon = 10^{-2}$ and $10^{-3}$, respectively, and $M_U \sim 0.1 - 1$~GeV. 

The above feature can be clearly seen in Fig. ~\ref{fig:fig2}, which shows the invariant mass distribution of the muon pairs for three values of the kinetic mixing parameter and $U$ boson mass at DA$\Phi$NE energies ($\sqrt{s} = 1.02$~GeV), as obtained through our calculation in the lowest order approximation. The contribution of the QED background processes is taken into account by means of the exact matrix elements for 
$e^+ e^-  \to \gamma^* \to l^+ l^- \gamma, l = e,\mu$ \footnote{For the electron final state, there is another relevant background 
which should be considered, namely $e^+ e^- \to \gamma\gamma$ with conversion of one of the two photons in the beam pipe or
the inner wall of the detector. This background would require appropriate studies because of its instrumental origin and can be, however, made negligible for small dark photon masses by requiring cuts on the reconstructed invariant mass and vertex of the pair \cite{Li:2009wz,Bossi:2009uw}.}
implemented in the current version of BabaYaga@NLO \cite{Balossini:2006wc,Balossini:2008xr}. 

At flavor factories multiple soft and collinear radiation emitted by the colliding beams may have a strong impact 
on the measured cross section and on the shape of the distributions. The effect of higher order corrections is taken into account using the popular QED structure function approach \cite{Kuraev:1985hb,Nicrosini:1986sm}. Initial state radiation (ISR) modifies the tree level cross section as follows
\begin{equation}
d\sigma (s,t) \, = \, \int_{0}^{1} \, dx_1 \, dx_2 \, d\sigma_{0} (\hat{s}, \hat{t}) \, D(x_1, \hat{s}) \, D(x_2, \hat{s})
\label{eq:ddin}
\end{equation}
where $\hat{s}, \hat{t}$ are the Mandelstam invariants after initial state photon radiation and the electron structure function $D(x,s)$ describes multiple soft and hard photon emission in the collinear approximation according to the expression 
\begin{eqnarray}
&&D(x,s) = \frac{\exp\left[ \frac{\beta}{2} (\frac{3}{4}-\gamma_E) \right]}{\Gamma(1+\frac{\beta}{2})}
\frac{\beta}{2} \, (1-x)^{\frac{\beta}{2}-1} \, - \, \frac{\beta}{4} (1+x) \nonumber\\
&& \!\!\!\!+\frac{\beta^2}{32} \left[ (1+x)(-4\ln(1-x)+3\ln x) \!-4 \frac{\ln x}{1-x} -\!5 -x \right]
\label{eq:dform}
\end{eqnarray}

where $\beta = 2\alpha/\pi \, \big( \log(s/m_e^2) - 1 \big)$ is the large collinear factor dependent on the radiating fermion mass and
$x$ the four-momentum fraction of the electron(positron) after photon shower. In Eq.~(\ref{eq:dform}) $\Gamma$ is the Euler gamma-function and $\gamma_E \approx 0.5772$ the Euler-Mascheroni constant. Because typical event selections are characterized by cuts on the kinematical variables of the observed particles,  the inclusive conditions  of the Kinoshita-Lee-Nauenberg (KLN) theorem for the cancellation of logarithmic enhancements are not fulfilled and final state radiation (FSR) must be taken into account as well. We implement it in our calculation by including two further structure functions in Eq.~(\ref{eq:ddin}) to describe photon emission from the final state leptons, 
according to the following equation:
\begin{eqnarray}
d\sigma (s,t) \, &=& \, \int_{0}^{1} \, dx_1 \, dx_2 \, dy_1 \, dy_2\, d\sigma_{0} (\hat{s}, \hat{t}) \, \nonumber \\
& &D^{e}(x_1, \hat{s}) \, D^{e}(x_2,\hat{s}) \,  D^{l}(y_1, s') \, D^{l}(y_2, s') .
\label{eq:infin}
\end{eqnarray}
In the equation above the four structure functions $D$ are given by the explicit expression of Eq.~(\ref{eq:dform}). In particular, $D^{e}(x_i, \hat{s})$ are the structure functions of initial state electrons giving the probability of finding, inside an electron, an electron with four-momentum fraction $x_i$ at a virtuality scale $\hat{s}$, $D^{l}(y_i, s')$ are the structure functions of the final-state leptons, computed with  
$\beta \equiv \beta_f = \beta_e, \beta_\mu$ for the electron and muon channel, respectively, and at the scale $s'$ given by the lepton pair invariant mass.  
Equation~(\ref{eq:infin}) takes into account any experimental cuts, that are implemented by means of a rejection algorithm on an event by event basis. More details on the approach can be found in Ref.~\cite{Montagna:1993te}.  

Being an effect of size comparable to that of photon radiation, we also consider in our calculation 
the running of the electromagnetic coupling constant in the hard scattering cross section according to (see Ref.~\cite{Actis:2010gg} and references therein)
\begin{eqnarray}
& &\alpha(q^2) = \frac{\alpha}{1- \Delta \alpha(q^2)} \nonumber \\
& &\Delta \alpha(q^2) = \Delta \alpha_l(q^2) + \Delta \alpha_h(q^2) 
\label{eq:vacpol}
\end{eqnarray}
where $\Delta \alpha_l(q^2)$ is the leptonic contribution to photonic vacuum polarization (analytically known) and  $\Delta \alpha_h(q^2)$ is the non-perturbative hadronic contribution,  included according to the parameterization of Ref.~\cite{Jegerlehner:2006ju}. The typical effect of vacuum polarization is to enhance the cross section of some per cent, as explicitly checked in the present calculation. 

In order to clarify the role of ISR and FSR in the peak region, we show in Fig. \ref{fig:fig3} 
the impact of ISR and FSR separately on the invariant mass distribution of the muon pair in a narrow region 
around $U$ boson resonant production. A $U$ boson mass of 0.98 GeV at $\sqrt{s} = 1.02$~GeV
is considered. Because the Breit-Wigner shape of the muon pair invariant mass is induced by the emission of a sufficiently
hard, detected photon (radiative return mechanism), it can be {\it a priori} expected that the
additional ISR is dominated around the peak by the emission of soft photons for a 
$U$ boson mass close to the available c.m. energy. Actually,  not to destroy the resonance,
ISR photons  emitted additionally to the radiative return photon must have an energy
fraction smaller than $1 - M_U/\sqrt{s} = \Delta E/E$, $\Delta E/E$ being $\ll 1$ if $M_U 
\simeq \sqrt{s}$. Therefore, ISR must have the typical soft photon effect of reducing the
peak cross section, as observed e.g. in $Z$ resonant production at LEP. This
effect is clearly visible in Fig. \ref{fig:fig3}. 
On the other hand, independently of the $U$ boson mass value w.r.t. the c.m. energy,
the main contribution of FSR is degrading the momenta of the final state leptons.
This reduces the fraction of events with invariant mass peaked at the $U$ boson mass
and enhances the number of signal events with invariant mass smaller than the resonance
value, giving rise to a tail on the left of the peak. These effects due to 
FSR can be clearly seen in Fig. \ref{fig:fig3}.

The separate contribution of ISR and FSR is also shown in Fig. \ref{fig:fig4}
for $M_U = 6 $~GeV at $\sqrt{s} = 10 $~GeV. The above reasoning about FSR still 
applies and a left-tailed broadening of the peak is again observed. On the 
contrary, because now $M_U$ is definitely smaller than the available c.m. energy,
no tight cut off is  present for the maximum energy carried away by the
ISR photons. Said differently, resonant production is in this case possible 
also in the presence of sufficiently hard IS photons whose effect is, through
convolution, enhancing the peak. This can be clearly observed in Fig. \ref{fig:fig4}.

More in general, we studied the impact of photon radiation both for the muon and electron channels, for different values of the $U$ boson mass and kinetic mixing parameter. An example of the effects  due to higher order QED corrections, as obtained through our generator, is given in Fig.~\ref{fig:fig3}, which shows the relative contribution of  the radiation on the invariant mass spectrum of the lepton pairs 
in the presence of standard selection cuts  at DA$\Phi$NE, $\sqrt{s} = 1.02$~GeV and taking also into account detector resolution effects (see the next section).  It can be seen that the differential cross section of  the muon final state receives a moderate positive correction, of some per cent, for a 
relatively small muon-pair invariant mass,  (much) smaller than the available centre of mass (c.m.) energy, while a larger negative correction, at the level of 10--20\%, shows up for $M_{\mu^+ \mu^-} \simeq \sqrt{s}$. For the electron channel the steeper behavior of the lowest order invariant mass distribution makes the corrections much larger, 
of the order of 
+20\% for small/intermediate invariant masses and of about -30--40\% for $M_{l^+ l^-} \simeq \sqrt{s}$. In the figure a $U$ boson mass of 0.98~GeV has been considered, 
but the shape of the correction is substantially independent of the choice of c.m. energy and $M_U$, and thus reflects a general situation. 
If one considers  the constraints on 
the $U$ boson mass coming from astrophysical observations, i.e. $M_U \leq 2 M_p$, $M_p$ being the proton mass, in order to avoid antiproton production through DM annihilation, the latter situation is of interest only  for searches at the 
$\Phi$-factory DA$\Phi$NE and can be explained in terms of the dominance of soft photon emission when $M_{l^+ l^-} \simeq \sqrt{s}$.  In a broader class of new physics models, this relevant effect shows up when the $U$ boson mass is close to the kinematical boundary imposed by the available c.m.  energy. 
As a whole, these results point out that the control of photon radiation effects is mandatory at flavor factories for a reliable simulation of all the distributions useful as discovery tools of a dark photon signature. 
\begin{figure}
\begin{center}
\resizebox{0.5\textwidth}{!}{%
\includegraphics{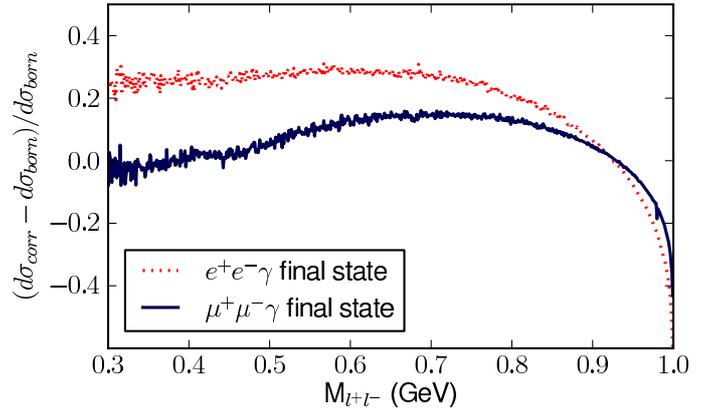}
}
\caption{Relative effect of photon radiation on the invariant mass distribution (for brevity $d \sigma$ in the plot) of the muon and electron pairs, for standard cuts at DA$\Phi$NE ($\sqrt{s} = 1.02$~GeV), and dark photon parameters $M_U = 0.98$~GeV, $\epsilon = 1 \times 10^{-3}$. }
\label{fig:fig5}
\end{center}
\end{figure}

\section{Experimental Sensitivity}
\label{sect3}
On the grounds of the calculation described in Section~\ref{sect2}, we revisited the experimental sensitivity to a dark force 
signal evaluated in the literature, and considered novel search strategies and 
event selections not addressed so far. For concreteness, 
we consider the case of the KLOE/KLOE-2 experiment at the upgraded DA$\Phi$NE~\cite{AmelinoCamelia:2010me}
and of present and future experiments
at $B$ and super-$B$ factories~\cite{Bona:2007qt,Golob:2010dm}. We assume an integrated luminosity $L$ of $5~{\rm fb}^{-1}$ for KLOE/KLOE-2
and $L = 500~{\rm fb}^{-1}, 100~{\rm ab}^{-1}$ for a $B$ and super-$B$ factory, respectively.
 We compute the statistical significance as
\begin{equation}
\frac{N_S}{\sqrt{N_B}} \, = \, \frac{L \, (\sigma_F - \sigma_B)}{\sqrt{L \sigma_B}}
\label{eq:statsig}
\end{equation}
requiring the above ratio to be greater than five for discovery. In Eq.~(\ref{eq:statsig}) $\sigma_F$ is the full cross section including the exchange of virtual photons and $U$ bosons, $\sigma_B$ the background cross section, $N_S$ and $N_B$ the expected number of signal and background events, respectively. To simulate detector acceptances, we also impose the following energy and angular cuts (from now on large angle selection)
\begin{eqnarray*}
&&{\rm KLOE} \quad \quad \, \, 35^\circ \leq \theta_{l^{\pm},\gamma} \leq 145^\circ \,\, E_{l^{\pm},\gamma} \geq 10~{\rm MeV} \\
&&B~{\rm factory} \quad 30^\circ \leq \theta_{l^{\pm},\gamma} \leq 150^\circ \,\, E_{\gamma,(l^{\pm})} \geq 20,(30)~{\rm MeV} 
\end{eqnarray*}
using as c.m. energies $\sqrt{s} = 1.02, 10.56$~GeV, respectively. Additionally, signal events can be detected as peaks in the lepton pair invariant mass close to the value $M_U$, as visible in Fig.~\ref{fig:fig2}, but only in a window $M_U \pm \delta_M$, where 
$\delta_M$ is a cut to reject backgrounds or the mass resolution of the detector. To avoid washing  out the signal against the background, 
$\delta_M$ should optimally coincide with the detector resolution, which is a crucial parameter for these searches. As detector resolutions we use $\delta_M = \pm 1$~MeV for KLOE/KLOE-2 and the values obtained according to the empirical relations of 
Ref. \cite{Reece:2009un} for $B$/super-$B$ factories, giving $\delta_M \sim \pm \, [1-10]$~MeV for a mass $M_U$ starting from  $0.1$~GeV and up to a few GeV. 

\begin{figure}
\begin{center}
\resizebox{0.5\textwidth}{!}{%
\includegraphics{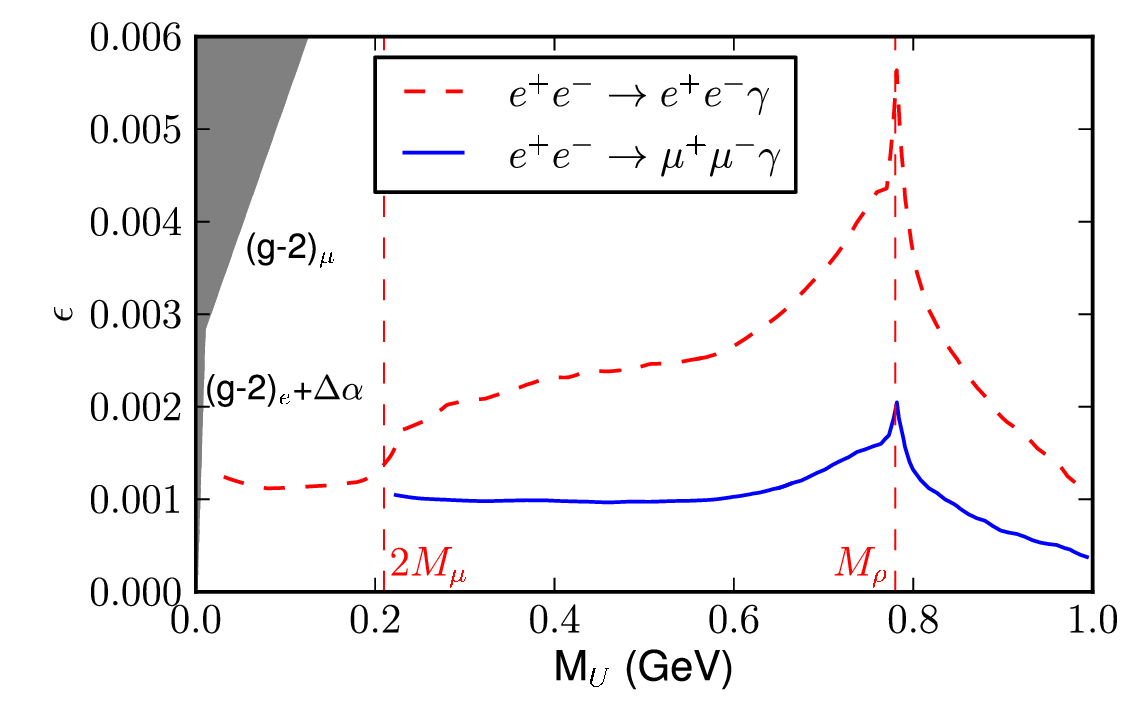}
}
\caption{Discovery potential, defined as the value $\epsilon$ at which $N_S/\sqrt{N_B}$ = 5, as a function of the dark 
photon mass $M_U$, at KLOE/KLOE-2 for an integrated luminosity $L = 5$~fb$^{-1}$ and large angle selection cuts. }
\label{fig:fig6}
\end{center}
\end{figure}

In Fig.~\ref{fig:fig6} we show the reach potential of KLOE/KLOE-2 experiment whose sensitivity has been discussed in the literature 
only in the exploratory work of Ref.~\cite{Bossi:2009uw}. We give results for the two leptonic final states 
according to our predictions in the lowest order approximation. Actually, we observed that  the signal and background cross sections are both affected by corrections of about the same amount; hence the systematics induced by photon radiation largely 
cancel in the experimental sensitivity 
and the conclusions in the presence of QED radiative corrections confirm the ones obtained in the lowest order approximation. 
The sensitivity to the kinetic mixing parameter 
$\epsilon$ is shown as a function of the $U$ boson mass. The grey areas correspond to the exclusion limits 
imposed by the precision measurements of the anomalous magnetic moment of the leptons and 
of $\alpha_{\rm QED}$ \cite{Pospelov:2008zw}.
As already emphasized in previous studies, the muon channel has better reach than the $e^+e^-$ channel which is affected by a large radiative Bhabha scattering background. Both channels have a
sensitivity which is significantly degraded if the $U$ boson mass is around the $\rho$ resonance because the branching 
fraction $U \to l^+ l^-$ is suppressed by the dominant decay mode $U \to \pi^+ \pi^-$.  Therefore, in the region $M_U$ around 
the mass of the $\rho$ it would be more convenient to utilize the $\pi^+\pi^-$ decay mode to recover the loss in
the reach due to suppression of the leptonic branching ratio. We note that the 
maximum sensitivity achievable at the upgraded DA$\Phi$NE with a luminosity $L \simeq 5$~fb$^{-1}$, i.e. $\epsilon \sim 0.001-0.002$, is equivalent, as we explicitly checked, to that of the present $B$-factory experiments BaBar/Belle with $L \simeq 500$~fb$^{-1}$~\cite{Reece:2009un}, since the reach on $\epsilon$ follows, from Eq.~(\ref{eq:statsig}), the rule $\epsilon^2 \propto (s/L)^{1/2}$. For the case already addressed in the literature of a $B$-factory with an integrated luminosity of 500~fb$^{-1}$ 
we compared our results for the reach potential with those derived in Ref. \cite{Reece:2009un}, observing 
very good agreement. This demonstrates that the sensitivity limits obtained considering a $U$ boson signal in the lowest order, on-shell approximation are robust and unaffected by a full signal calculation, since, as argued in Ref. \cite{Reece:2009un}, the interference between the resonance and continuum background is negligible because of the extremely small width of the $U$ boson. Furthermore, the 
sensitivity is not significantly affected by the contribution of QED corrections, especially in the muon channel for small or intermediate 
values of the invariant mass, as 
can be qualitatively understood as follows. Because the tree level and QED corrected signal and background cross sections 
receive relative corrections $\delta$ of about the same size, when 
considering realistic detector resolution values that tend to
reduce the impact of the radiation on the signal cross section, the following relations hold
\begin{equation}
\sigma_S^{\rm QED} \, \simeq (1 + \delta) \, \sigma_S^{0}   \qquad \sigma_B^{\rm QED} \simeq (1+\delta) \sigma_B^{0} 
\label{eq:sigmacorr}
\end{equation}
where, as above, the signal cross section $\sigma_S$ is defined as the difference between the full and background cross section. 
It then follows from Eq.~(\ref{eq:statsig}) that the maximum sensitivity achievable according to a lowest order prediction is modified by the effect of photon radiation as 
\begin{equation}
\epsilon^{0} \to \epsilon^{\rm QED} \equiv \tilde{\epsilon}^{0} \simeq \sqrt{1+ \delta/2} \, \epsilon^{0}
\label{eq:epscorr}
\end{equation}
that explains the results obtained in our simulations. 
This {\it a posteriori} reinforces the robustness of the lowest order sensitivity results, showing that they are stable against QED radiative corrections. 
Note that, according to Eq.~(\ref{eq:epscorr}) and the numerical results shown in Fig.~\ref{fig:fig3}, the 
experimental sensitivity is deteriorated by QED corrections when multiple soft photon emission dominates, like for 
$M_U \simeq \sqrt{s}$, whereas it slightly improves, especially in the electron channel, for small and intermediate $U$ boson mass values, because of the positive correction induced by hard bremsstrahlung. 

\begin{figure}
\begin{center}
\resizebox{0.5\textwidth}{!}{%
\includegraphics{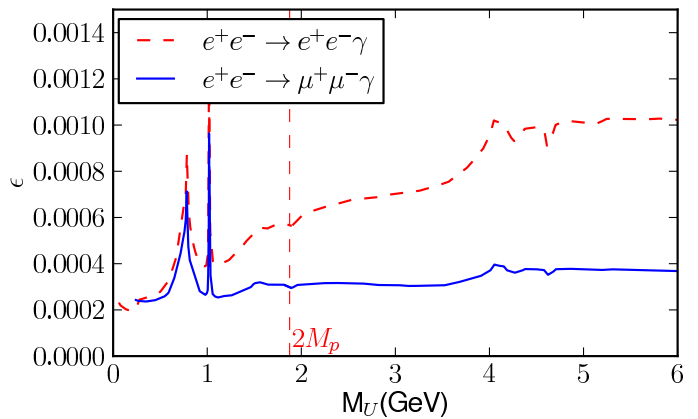}
}
\caption{The same as~\ref{fig:fig6} at a super-$B$ factory with an assumed luminosity $L = 100$~ab$^{-1}$.}
\label{fig:fig7}
\end{center}
\end{figure}

In Fig.~\ref{fig:fig7} we present the discovery potential at a super-$B$ factory, a situation not yet explored in the literature, 
showing again the results for the two lepton final states. The muon channel has a definitely better reach than the electron one, like at DA$\Phi$NE. The sensitivity that can be reached at a super-$B$ is well below $10^{-3}$ for all 
mass values, with the exception of the region $M_U \sim M_\phi$. 
However, the latter situation corresponds to a very narrow interval because of the extremely small value of the branching ratios of the 
$\phi$ resonance into leptons. The same holds true for all the other narrow hadronic resonances,  not shown in the figure because of their tiny width. 
Hence the large data set which can be in principle collected at a super-$B$ collider will allow to probe values of the kinetic mixing parameter about an order of magnitude smaller than those reachable by present flavor factories~\footnote{We notice in passing that the sensitivity in the electron channel could be significantly improved by adopting asymmetric angular cuts for the electrons, namely by cutting the forward peaking behavior due to $t$-channel photon exchange, thus increasing the signal/background ratio. Nevertheless, the muon channel remains the most performing in the large angle selection. }. 

\begin{figure}
\begin{center}
\resizebox{0.5\textwidth}{!}{%
\includegraphics{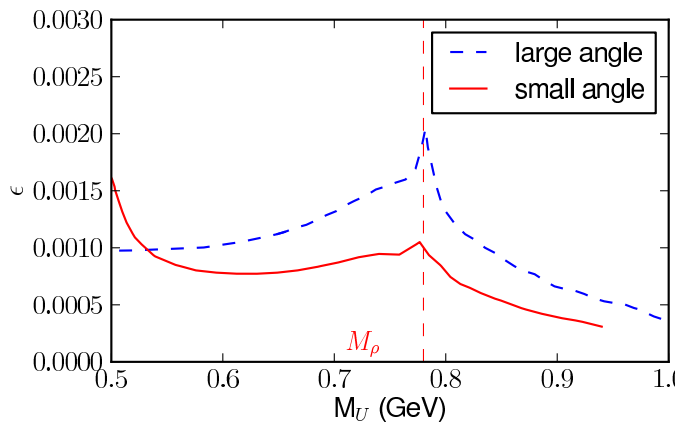}
}
\caption{Discovery potential for the $\mu^+ \mu^- \gamma$ channel as a function of the dark 
photon mass $M_U$, at KLOE/KLOE-2 for an integrated luminosity $L = 5$~fb$^{-1}$. The sensitivity according to the small angle selection cuts as in the text is shown in comparison with the large angle selection. }
\label{fig:fig8}
\end{center}
\end{figure}

\begin{figure}
\begin{center}
\resizebox{0.5\textwidth}{!}{%
\includegraphics{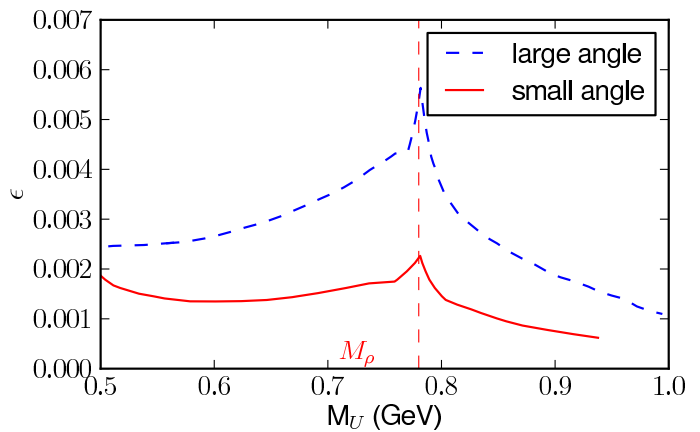}
}
\caption{The same as~\ref{fig:fig8} for the electron channel.}
\label{fig:fig9}
\end{center}
\end{figure}

In Figs.~\ref{fig:fig8} and \ref{fig:fig9} the sensitivity for the cuts defined as (from now on small angle selection)
\begin{eqnarray*}
{\rm KLOE} \quad \quad \, \, &&35^\circ \leq \theta_{l^{\pm}} \leq 145^\circ,  \,\, E_{l^{\pm}} \geq 10~{\rm MeV} \\
&&\vert \cos(\theta_\gamma) \vert \geq \cos(15^\circ) 
\end{eqnarray*}
is shown in comparison with the sensitivity for the large angle selection previously discussed. 
This event selection is actually used at DA$\Phi$NE for the measurement of the $\pi \pi \gamma $ cross section~\cite{Actis:2010gg} and has been never considered before for this kind of studies. The results are shown for lepton pairs invariant masses larger than 0.5~GeV, since below this value the cross sections drop steeply to very small values. 
As can be noticed, in both the electron and muon channel there is a relevant gain in sensitivity,   making this selection a very interesting research tool.

\section{Summary}
\label{sect4}
To summarize, in the present paper the search for a light and weakly coupled new gauge boson at contemporary and future flavor factories has been analyzed. 
After  performing an exact tree level calculation of the signal and background processes contributing to the signature  $e^+ e^-  \to \mu^+ \mu^- \gamma$,  $e^+ e^- \gamma$, 
we computed the effect of the most important higher order corrections induced by multiple photon radiation and vacuum polarization, that are a source of relevant systematic effects. Next, we used our calculation in order to assess its impact on the experimental sensitivity to the dark photon parameters as evaluated in the literature, and showed how this can be enhanced by means of event selections not considered so far. 

More in detail, we have shown that the effect of QED  radiative corrections can be of the order of tens of per cent on the invariant mass distributions and, more in general, non-negligible in the computation of the various observables of physical interest. Therefore, in a sensible data analysis, these corrections cannot be ignored. 

Using our calculation we have revisited the experimental sensitivity to a dark photon signal at present and future 
high-luminosity flavor factories. Moreover we considered also new event selections, not explored so far. 

As far as the first item is concerned, 
we have shown that the reach potential is only slightly affected by the inclusion of photon radiation effects, and thus {\it a posteriori} it has been shown to be robust with respect to the inclusion of potentially dangerous effects. 
We have shown that the maximum sensitivity on the kinetic mixing parameter is achievable through the study of the muon channel for all values of the $U$ boson mass above the muon pair production threshold, while the electron channel can allow to probe the $U$ boson mass region below it with comparable potential, provided the background 
$e^+ e^- \to \gamma\gamma$, with conversion of one photon into an electron pair, is appropriately rejected. We have analyzed two 
interesting cases not yet addressed in the literature corresponding to the luminosity achievable at KLOE/KLOE-2 and a future 
super-$B$ factory. The statistical significance to the kinetic mixing parameter is of the order of $\epsilon \sim 0.001$ at  KLOE/KLOE-2 and present $B$-factories and at the level of a few units in $10^{-4}$ at a very high luminosity super-$B$. 

As far as the second item is concerned, we considered a small angle event selection, similar to the one already used at DA$\Phi$NE for the measurement of the $\pi \pi \gamma$ cross section, finding that this allows to obtain a considerable enhancement in the sensitivity for a $U$ boson mass above, say, 0.5~GeV, both for the electron and muon channels. 

Our calculation is available in an improved version of the generator BabaYaga@NLO, which can be used for event simulations and data analysis in these new physics searches~\cite{pvweb}.

\begin{acknowledgement}
We are grateful to Fabio Bossi of KLOE Collaboration for bringing 
our attention to the search for dark force signals at flavor factories 
and helpful correspondence. We also thank Federico Nguyen of KLOE and 
Mauro Moretti for useful discussions, as well as  Graziano Venanzoni 
of KLOE Collaboration  for helpful suggestions. We are indebted with 
Thomas Teubner for providing us with the routine for the calculation 
of the $R$ ratio.
\end{acknowledgement}
\bibliographystyle{epjc}
\bibliography{dark-force}

\end{document}